# Martensitic transformation in zirconia.

# Part I: nanometer scale prediction and measurement of transformation induced relief

Sylvain Deville[†], Gérard Guénin, and Jérôme Chevalier

Associate Research Unit 5510, Materials Science Department, National Institute of Applied Science (GEMPPM-INSA), Bat. B. Pascal, 20 av. A. Einstein, 69621 Villeurbanne Cedex, FRANCE

**Abstract**

We investigate by atomic force microscopy (AFM) the surface relief resulting from martensitic tetragonal to monoclinic phase transformation induced by low temperature autoclave aging in ceria-stabilized zirconia. AFM appears as a very powerful tool to investigate martensite relief quantitatively and with a great precision. The crystallographic phenomenological theory is used to predict the expected relief induced by the transformation, for the particular case of lattice correspondence ABC1, where tetragonal c axis becomes the monoclinic c axis. A model for variants spatial arrangement for this lattice correspondence is proposed and validated by the experimental observations. An excellent agreement is found between the quantitative calculations outputs and the experimental measurements at nanometer scale yielded by AFM. All the observed features are explained fully quantitatively by the calculations, with discrepancies between calculations and quantitative experimental measurements within the measurements and calculations precision range. In particular, the crystallographic orientation of the transformed grains is determined from the local characteristics of transformation induced relief. It is finally demonstrated that the strain energy is the controlling factor of the surface transformation induced by low temperature autoclave treatments in this material.

**Résumé**

Nous avons étudié par microscopie à force atomique (MFA) les caractéristiques du relief de surface induit par la transformation martensitique quadratique-monoclinique de la zircone cériée. La transformation a été provoquée par des traitements de vieillissement en autoclave. La MFA apparaît comme une technique extrêmement puissante pour étudié le relief martensitique de manière quantitative et avec une grande précision. La théorie phénoménologique cristallographique a été appliquée pour prédire le relief de surface induit par la transformation, dans le cas particulier de la correspondance ABC1, ou l'axe quadratique c devient l'axe monoclinique c. Un modèle pour l'arrangement spatial des variantes pour cette correspondance de réseau est proposé, et validé par les observations expérimentales. Une excellente concordance est obtenue entre les résultats quantitatifs des calculs et les mesures expérimentales à l'échelle nanométrique, obtenues par MFA. Toutes les caractéristiques du relief de surface observées peuvent être interprétées par les calculs, les différences observées restant dans les marges d'erreur des observations et la précision des calculs. En particulier, l'orientation cristallographique des grains est déterminée à partir des caractéristiques locales du relief induit par la transformation. Finalement, nous démontrons que l'énergie de déformation est le facteur contrôlant la transformation de surface induite par les traitements en autoclave pour ce matériau.

## 1. Introduction

The martensite transformation model, developed initially by Bain [1], has now been the object of almost a century of investigations. It was originally associated to the transformation occurring in steel quenching: the austenite, formed at high temperature, transforms to martensite by rapid cooling to avoid species diffusion. By extension, the martensitic transformation term is associated to transitions occurring in a fairly large number of materials exhibiting characteristics described by the following definition: "A martensitic transformation is a first order displacive structural transition exhibiting lattice invariant strain, essentially composed of a shear strain." The first order displacive term means that the transformation involves atomic displacements small but finite (about a tenth of the atomic spacing), and perfectly correlated for a large number of atoms. No atomic diffusion is occurring during the transition, so that it may occur at any temperature, without changing either the atomic order or the chemical composition. The atomic shifts are such that they lead to a homogeneous strain of the crystal lattice, with a small volume change as compared to the shear components.

The models developed to describe a martensitic transformation are however not specific to a particular material. Martensitic transformation also occurs in a number of minerals and ceramics, which have also been investigated for decades. Among these ceramics, the tetragonal to monoclinic phase transformation in zirconia is martensitic in nature.

This transformation has been investigated for both its positive (transformation toughening) and negative consequences (low temperature degradation, microcracking). The tetragonal ($P4_2/nmc$) to monoclinic ($P2_1/c$) transformation occurs at ~1223K ($M_s$) on cooling in pure zirconia, and is accompanied by a shear strain of ~0.16 and a volume expansion of ~0.04. The transformation is reversible and occurs at ~1423K ($A_s$) on heating. By alloying with oxides such as $Y_2O_3$ or $CeO_2$, a fully tetragonal phase microstructure is obtained, with an $M_s$ temperature such that spontaneous transformation does not occur on cooling to room temperature. Transformation can then be triggered by the action of water, by low temperature treatments (typically 350-500K) in steam autoclave.

The macroscopic consequences of the transformation have been widely investigated, and several phenomenological theories have been developed, based on the crystallographic characteristics, to predict in particular the expected relief changes associated to the transformation strain. However, very few quantitative reports of the transformation induced relief may be found in the literature, which is not surprising considering the scale at which the transformation is occurring. The martensite transformation has been mainly investigated in metals, for its importance in shape memory alloys. In absence of an appropriate technique allowing investigating the relief at the nanometer scale, the transformation has been studied mainly by optical methods or with transmission electron microscopy. The modified mechanical and chemical environments in thin foils affect drastically the transformation behaviour. These results might therefore be questioned when compared to the behaviour of bulk samples.

Quite fortunately, the recent developments of scanning tunnelling microscopy (STM) and atomic force microscopy (AFM) [2] provide the scientific community with powerful tools to investigate phenomena characterized by relief variations at a nanometer scale and even below. Few reports might be found on martensite in iron based alloys [3-5]. The absence of specific sample preparation and the possibility of observing bulk non-conductive materials make AFM very attractive to study martensitic transformation in ceramics and zirconia in particular. Very few reports [6-8] can be found of martensitic relief investigations in zirconia by atomic force microscopy, all of which being still very preliminary.

In the present paper, relief induced by the martensitic transformation in ceria stabilized zirconia is predicted and measured with precision. The phenomenological crystallographic theory is applied with Kajiwara's approach [9] to this particular case and the experimental results are compared to the calculations outputs. A model for variants growth and spatial arrangement is proposed and validated by the experimental observations. All of the complex spatial arrangements observed experimentally are completely explained in a quantitative manner by calculation results.

## 2. Experimental techniques

Ceria stabilized zirconia (Ce-TZP) materials were processed by classical processing route, using Zirconia Sales Ltd powders, with isostatic pressing and sintering at 1823 K for two hours. Residual porosity was negligible. Samples were mirror polished with standard diamond based products.

The martensitic transformation was induced by a thermal treatment in water vapour. This kind of treatment is known to induce the tetragonal to monoclinic phase transformation at the surface of zirconia [10,11], though the underlying chemical mechanism for transformation is still a matter of debate. These treatments were conducted in autoclave at 413 K, in saturated water vapour atmosphere, with a 2 bar pressure, inducing phase transformation at the surface of the samples with time.

AFM experiments were carried out with a D3100 nanoscope from *Digital Instruments Inc.*, using oxide sharpened silicon nitride probes (*Nanosensor*, *CONT-R* model) in contact mode, with an average scanning speed of 10 µm.s$^{-1}$. Since the t-m phase transformation is accompanied by a large strain, surface relief is modified by the formation of monoclinic phase. The vertical resolution of AFM allows following very precisely the transformation induced relief.

## 3. Correspondence choice

In zirconia, it has been shown [12] that several lattice correspondences and lattice invariant shears can be followed. In zirconia, three possible lattice correspondences exist, called ABC, BCA and CAB, which imply that $a_t$, $b_t$, $c_t$ of the tetragonal phase changes into $a_m$, $b_m$, $c_m$; $b_m$, $c_m$, $a_m$ and $c_m$, $a_m$, $b_m$ respectively. The correspondence choice for applying here the crystallographic phenomenological theory [13,14] is made in regards of the first experimental features revealed by AFM, shown in Figure 1:

- Transformation induced relief is locally constituted by sets of 4 variants, which seem to exhibit a fourfold symmetry. This indicates that the free surfaces where the observations are done are not very far from the $(001)_t$ plane of the tetragonal phase.

- The relief displacement occurs always outside of the matrix surface. This suggests that the observed variants are the ones which reject at most the volume change towards the surface. This should lead to a large decrease of the transformation induced elastic stresses within the volume, making this configuration much more favourable to occur from an energetic point of view.

Therefore we are looking for compatible four variants with fourfold symmetry which give common transformation displacements essentially directed towards the outside of the free surface which is not far from $(001)_t$. The main features of the martensite plates associated to the three correspondences and calculated from the phenomenological theory are given in Tab. 1; details of the present calculations will be given in another paper. It is nonetheless worth mentioning the results obtained here are exactly the same than those reported by Kelly and Rose, though the method used for calculations is different. From each correspondence two kinds of variants can be observed, which is seen with more clarity if the components of habit plane normals (*hpn*) and shape strain (*ss*) below 6/1000 are neglected (Tab. 2)

It has be shown that self accommodating pairs of variants can be built from these elementary variants with junction planes being simple $(100)_t$, $(010)_t$ or $(001)_t$ planes. Fig. 2 shows schematically these pairs and the pairs deduced from fourfold symmetry, with their effect on the free surface when this one is perpendicular to the junction plane. From Fig. 2 and Tab. 2, it can be seen that only the two situations ABC1 and BCA2 can provide the expected group of four variants. Indeed, in these cases, the variants have shape strain directions with a common large component along the $c_t$ axis and a common free surface. The BCA2 four variants group has the following approximate characteristics:

$$HP_1 = \begin{bmatrix} 0 \\ -0.99 \\ -0.03 \end{bmatrix} \quad SS_1 = \begin{bmatrix} 0 \\ -0.056 \\ 0.167 \end{bmatrix} \quad HP_2 = \begin{bmatrix} 0 \\ 0.99 \\ -0.03 \end{bmatrix} \quad SS_2 = \begin{bmatrix} 0 \\ 0.056 \\ 0.167 \end{bmatrix}$$

$$HP_3 = \begin{bmatrix} -0.99 \\ 0 \\ -0.03 \end{bmatrix} \quad SS_3 = \begin{bmatrix} -0.056 \\ 0 \\ 0.167 \end{bmatrix} \quad HP_4 = \begin{bmatrix} 0.99 \\ 0 \\ -0.03 \end{bmatrix} \quad SS_4 = \begin{bmatrix} 0.056 \\ 0 \\ 0.167 \end{bmatrix}$$

The behaviour between variants 1 and 2 close to the surface is displayed in Fig. 3: to have the common component of the displacement going out of the free surface, the variant thickness must slightly increase when going inside the crystal, which makes impossible the ends of these variants inside the crystal without strain incompatibilities. Moreover a quite large component of the displacement is perpendicular to the junction plane (and to the habit planes), implying large compression internal strains, even close to the surface. This situation is therefore very unlikely to occur and is rejected. This choice is comforted by the absence of experimental observations in the literature of the BCA correspondence in the ceria stabilized zirconia.

From the phenomenological theory, the ABC1 solution has the following accurate characteristics expressed in the orthonormal axis system bounded to the tetragonal lattice system:

$$M_{pt} = \begin{bmatrix} 1.00248 & -0.00001 & -0.00078 \\ -0.00268 & 1.00002 & 0.00085 \\ -.15638 & 0.00089 & 1.04928 \end{bmatrix} \quad HP_{pt} = \begin{bmatrix} -0.9538 \\ 0.0055 \\ 0.3006 \end{bmatrix} \quad SS_{pt} = \begin{bmatrix} -0.0026 \\ 0.0028 \\ 0.1640 \end{bmatrix}$$

$M_{pt}$ is the matrix transforming a vector of the parent phase into the same vector in the product phase, $HP_{pt}$ is the habit plane normal and $SS_{pt}$ is the shape strain vector. Neglecting the components below 6/1000, the ABC1 four variants have the following characteristics (the shape strain vector is the same for the four variants):

$$M_1 = \begin{bmatrix} 1 & 0 & 0 \\ 0 & 1 & 0 \\ 0 & -0.156 & 1.049 \end{bmatrix} \quad HP_1 = \begin{bmatrix} 0 \\ -0.954 \\ 0.301 \end{bmatrix} \quad SS = \begin{bmatrix} 0 \\ 0 \\ 0.164 \end{bmatrix}$$

$$M_2 = \begin{bmatrix} 1 & 0 & 0 \\ 0 & 1 & 0 \\ 0 & 0.156 & 1.049 \end{bmatrix} \quad HP_2 = \begin{bmatrix} 0 \\ 0.954 \\ 0.301 \end{bmatrix}$$

$$M_3 = \begin{bmatrix} 1 & 0 & 0 \\ 0 & 1 & 0 \\ -0.156 & 0 & 1.049 \end{bmatrix} \quad HP_3 = \begin{bmatrix} -0.954 \\ 0 \\ 0.301 \end{bmatrix}$$

$$M_4 = \begin{bmatrix} 1 & 0 & 0 \\ 0 & 1 & 0 \\ 0.156 & 0 & 1.049 \end{bmatrix} \quad HP_4 = \begin{bmatrix} 0.954 \\ 0 \\ 0.301 \end{bmatrix}$$

A simple arrangement of these four variants when the free surface is $(001)_t$ is displayed in Fig. 4. All the interfaces matrix-martensite are habit planes and are unchanged during the transformation. All the interfaces between the variants are of the kind $\{100\}_t$ or $\{110\}_t$ and are jointly elongated in the direction $[001]_t$. Indeed, in this arrangement, every point of these interfaces is equidistant from the two involved variant habit planes and therefore moves by the same amount. As a consequence, all the volume change is exactly relaxed outside of the free surface and no internal stresses at all are generated inside the volume. In fact, as the strain directions are not exactly $[001]_t$ (Tab. 1), some internal strains are expected to remain. This kind of self accommodated group is therefore very likely to occur and is the good candidate for explaining the present observations.

It finally worth noticing the proposed arrangement is very similar to these observed in previous studies by TEM on thin foils [15-17], where $\{100\}_t$, $\{010\}_t$ and $\{110\}_t$ type junction planes where observed, along with habit planes of $\{103\}_t$ type and those derived from the fourfold symmetry. However, in the mentioned references, this arrangement was thought to be a feature specific of TEM thin foils. The very low thickness of thin foils does not lead to the same restrictions for transformation strain accommodation in the volume. The observed habit planes were however very close from the ones calculated here.

## 4. Transformation induced relief calculation

### 4.1 Ideal case

The spatial arrangement of Fig. 4 can be detailed in 2D (Fig. 5) to precise the surface angles produced by two opposite variants. On this figure, the displacement OO' common to variants 1 and 2 is given by 0.164 x $U_0O$ and/or 0.164 x $V_0O$. The displacement linearly decreases from points O to U (O to V) and from points O to I. The angle between the free surfaces modified by the two variants can easily be deduced (17.7°). AFM observations (e.g. Fig. 6) are close from expected by this analysis, but never quite exactly. This is quite logical considering that having the $c_t$ axis exactly perpendicular to the surface is a very particular case. In a general case the $c_t$ axis could be away from the perpendicular to the surface, the intersection of the variants group with the free surface is no more a rectangle, and the angles between the traces of the variant interfaces no more 90° or 45°. Moreover, angles of the free surfaces modified by the variants are no more 17.7°. This can be schematically shown in Fig. 5 where a free surface have been drawn away from $(001)_t$; it is worth noticing the displacement OO' is not modified by this operation.

### 4.2 Prediction of the relief as a function of the free surface axis inclination in relation to $c_t$

AFM experimental observations perpendicular to the surface can always be reduced as shown on the lower schemes (Fig. 6) from ABCD O'$_1$O'$_2$ to A'B'CDO'$_1$. This reduction corresponds to the 3D picture of fig. 7 (top left) when the free surface is $(001)_t$. The four lines IA, IB, IC and ID with IO are the limits of the four variants, the surface relief is characterized by the five points A, B, C, D, O'; the displacement is completely defined if IO is given, for example with IO = 1, OO' = 0.301 x 0.164 = 0.049. Let us now have a free surface axis inclined in relation to $c_t$ but passing always through the point O. The lines IA, IB, IC and ID remain the same but the intersections with the free surface are now A', B' C' and D'; O' is not changed (fig. 7, top right). These points characterise the new surface relief. The free surface can be defined by the two angles γ and δ as shown in Fig. 7 (bottom). A Matcad calculation and 3D representation have been built to display the set of four variants in 3D as a function γ and δ. From the view perpendicular to the free surface, it is therefore possible to compare with the corresponding AFM experimental picture. The best adjust gives $γ_o$ and $δ_o$ which define the free surface in the tetragonal axis system. This orientation must be coherent with the angles between the four faces of the relief. These angles can be calculated by using the transformation matrices of the four variants given above. Unfortunately, until now, no direct experimental surface orientation has succeeded to compare with the calculated ones, this point remaining a challenge for the future. Preliminary experiments by using electron backscattered diffraction (EBSD) are under consideration. The features of the microstructure (small grain size) and the technique in itself do not allow straightforward observations; this work is still under progress.

## 4.3 Comparison of the calculations and the experimental observations

Fig. 8 shows the comparison of the calculations with the experimental observations for three different cases. The Matcad calculations provide a 3D plot of the variants with both the traces at the free surface and the habit plans in the volume, like in Fig. 7. In a first look, all the arrangements seem very different. When reduced to their simple trace, where the four junction planes of $\{110\}_t$ type intersect at a common point (point $O'_1$ of Fig. 6), it appears that the two last cases are not so different. The extracted traces are used for fitting the model parameters γ and δ to the experimental observations. It is worth mentioning the fit is done as a function of the concordance of both the angles between the four faces of the relief and the habit and junction planes traces at the free surface. This adjustment provides the values of the angles relationships between the opposed faces of the relief, i.e. $α_{12}$ and $α_{34}$, which can be compared with the experimental measurements. It is also worth noticing the overall shape of the trace is much more sensitive to input parameters variations than the angles between the faces of the relief.

## 5. Influence of parameters variability

### 5.1 Lattice parameters

The lattice parameters were not measured directly in this study. Values from the literature [18] are used as input parameters for the calculations. To ensure this choice was not arbitrary, simulations were run with different lattice parameters [19,20] ($a_t = 0.5125$ nm, $c_t = 0.5220$, $a_m = 0.51928$, $b_m = 0.52043$, $c_m = 0.53617$, $\beta = 98.8°$). The new values of the **M** and **HP** are respectively (e.g. for variant 3):

$$M_3 = \begin{bmatrix} 1 & 0 & 0 \\ 0 & 1 & 0 \\ -0.155 & 0 & 1.043 \end{bmatrix} \qquad HP_3 = \begin{bmatrix} 0.963 \\ 0 \\ -0.268 \end{bmatrix}$$

By comparison with the previous results, noticeable differences are observed (e.g. volume increase decreased from 0.049 to 0.043), in particular in the habit planes indices. However, if the new values are used as input parameters in the developed model, very small variations on the predicted angles are obtained and the traces are nearly identical. The effect of the lattice parameters variations was investigated for a wide range of configurations, and resulting angle variations are found to be less than ±0.1° (typically ±0.05°). These variations are well within the experimental range of error, and thus do not affect significantly the results presented here.

### 5.2 Surface preparation

The quality of the surface finish is obviously of prime importance if highly reliable measurements are needed. Since very low angles variations (typically 0.2° to 0.5°) can be measured by AFM, it is important obtaining a surface as flat as possible. Fortunately, it is possible reaching excellent surface state with conventional polishing techniques in ceramics using diamond based products. Surface roughness's ($R_a$) of 2 nm are easily reached. The angle variations on the untransformed parts of the surroundings of relief features were systematically verified and found less than 0.5°, using analysis surfaces of 500 nm×500 nm. The typical depth of the observed polishing scratches is 1 to 2 nm. Considering the typical relief height increase (30 to 100 nm), it can be considered there is no need for reaching better surface finish.

## 5.3 Scanning direction

The influence of scanning direction was investigated, so as to assess the experimental expected precision. The same relief feature was imaged with three different scanning directions (Fig 9), i.e. -45°, 0° and 45°. The shapes resulting from the intersection of the habit planes and the free surface are extracted from the AFM images and compared to each other. When aligned along (CD) and on point D, a small rotation of the shape is observed. However, if now the various shapes are aligned so as to obtain the best fit, it can be seen that three of the sides are very well superimposed. An influence of the scanning direction on (CD) side is observed, with the (ADC) angle exhibiting a 10° variation range. If this value seems to be important, it is worth pointing out here the fit of the model is made as a function of all the traces of surface and relief and also the surface angles relationships. The range of error obtained after all the comparison is therefore in the range of order of 1° and not 10°. This behaviour can be related to the shape of the AFM probes. *Nanosensor*™ silicon probes (CONT-R type) were used in this study. These probes exhibit half cone angles of 20° to 25° along the direction of the cantilever axis, and 25° to 30° when seen from the side. An SEM image illustrating the anisotropy of the probe shape is shown in Fig 10. Depending on the scanning direction, the variation of the half cone angle of the side in contact with the relief will affect the images relief features by a simple convolution effect. Moreover, the probes are degraded fairly rapidly under use, so that the tip radius, originally announced at 10 nm, increased up to much larger values, increasing the convolution effect.

## 6. Complex features prediction and interpretation

The basic spatial arrangement described previously can be successfully applied for explaining more complex surface features observed experimentally.

## 6.1 Partial transformation

The first derivate arrangement observed by AFM is illustrated in Fig. 11. This arrangement presents a fourfold symmetry, with an untransformed part in the middle. The ABC1 correspondence characteristics can be used to describe the spatial accommodation and variants arrangement below the surface. This interpretation is plotted both in 3D and 2D in Fig. 11 for more clarity. From the 2D plot, it can be observed all the transformation strain is accommodated. The two variants $V_1$ and $V_2$ present a common junction plane similar to the fully transformed arrangement described previously. $V_3$ and $V_4$ are in a similar configuration, partially transformed. The particularity of this configuration is the presence of a residual untransformed volume of material, which is surrounded by the four variants. The interfaces between this portion and the variants are all habit planes described previously. The displacement of the opposed variant along the habit plane is constant, so that no residual transformation induced stresses should be expected in this configuration. This accommodation explains why such a partial transformation is possible.

## 6.2 Secondary variants and variants impingement

So-called secondary variants and variants impingement was also observed experimentally, as shown in Fig. 12. These secondary smaller variants pairs reflect an evolution of the transformation penetration depth, as sketched in Fig. 12. All the transformation strains are fully accommodated by this arrangement. It is also worth noticing that secondary variants are not bounded to the position of the main variants pair's junction plane. The secondary variants can be translated along the $a_t$ and $b_t$ axes, without modifying the transformation strain accommodation. When the translation along the b axis for example is large enough, it will result in an apparent impingement of variants groups, shown in Fig. 13. For this last configuration, the origins of the two groups are variants are two parallel segments below the surface.

## 6.3 L–shaped variants

When the three junction planes $\{100\}_t$, $\{010\}_t$ and $\{110\}_t$ intersect along a common line, L-shaped arrangements described in Fig. 14 are observed. Again, all the transformation strains are fully accommodated by this arrangement. The overall shape is also modified when the $c_t$ axis and the free surface normal are not exactly parallel. It is worth noticing the variants find their origin in the volume along an L-shape segment, which reflects exactly the surface arrangement.

Any more complex arrangement of surface relief presenting a fourfold symmetry can be interpreted by a combination of the arrangements described here, and taking into account the inclination of the $c_t$ axis with respect to the free surface.

## 7. Discussion

A statistical analysis of the average orientation relationship of the $c_t$ axis of the transformed grains to the surface was performed, using the analysis developed here. In this study, the martensitic transformation is triggered by treatments in water vapour autoclave. In this case, the transformation is initiated spontaneously on the more favourable sites from an energetic point of view. An example of the partially transformed surface in the first stage of the aging treatment is shown in Fig. 1. It is clearly observed the only transformed zones present all the same characteristics, i.e. their $c_t$ axis is nearly normal to the free surface. Of all the analysed transformed grains, the maximum orientation deviation of the $c_t$ axis to the surface ($\gamma$) is 30°. Factors affecting the transformation behaviour of zirconia based ceramics have now been widely investigated and documented. In some cases, the nucleation conditions [21] have been proposed for being the step controlling factor of transformation. Residual stresses could also be of prime importance. From the evidences provided here, it is quite clear that for the particular case of 10 mol.% $CeO_2$-TZP and also probably for different compositions (considering the very small variations of lattice parameters with the oxide content [22]) the factor controlling the transformation is the degree of strain accommodation. If the $c_t$ axis is nearly perpendicular to the surface, the transformation can proceed in the way described here, so that all the transformation strain can be accommodated vertically by free surface relief increase. However, if the transformation strain cannot be accommodated in the vertical direction, very large transformation induced stresses are expected, making the transformation much less favourable to occur, as observed experimentally.

Finally, it is worth noticing there are no firm reasons for this behaviour being the same when the stabilizing oxide is different. The habit planes and transformation strains have proven to be very sensitive to the lattice parameters. For yttria stabilised zirconia, for which the transformation behaviour has been widely documented, the lattice parameters are significantly different; fourfold symmetry observations have however been reported [8]. Such a high degree of transformation strain accommodation is not expected, so that other factors may control the transformation sensitivity, like stabilizing oxide distribution, grain size or residual stresses.

## 8. Conclusions

- Martensitic transformation induced relief in ceria stabilized zirconia was investigated by atomic force microscopy. The crystallographic phenomenological theory was applied and a model for variants spatial arrangement is proposed, based on crystallographic arguments, for lattice correspondence ABC1 where $c_t$ axis becomes $c_m$ axis. Volume change is exactly relaxed outside of the free surface and no internal stresses at all are generated inside the volume.

- Numerical calculations were conducted to predict transformation induced relief with surface relief angles relationships. Calculation results are found in excellent agreement with the experimental observations and measurements of nanometer scale transformation induced relief. Particular attention was paid to the limits of calculations and experimental measurements precision.

- This analysis was used to describe quantitatively all the relief features related to the correspondence ABC1, with more complex spatial arrangement. It is finally demonstrated that for 10 mol.% $CeO_2$-TZP, the degree of transformation strain accommodation is the controlling factor of the transformation, when martensite formation is triggered by hydrothermal treatment in water vapor autoclave.

- It is finally worth mentioning the experimental procedure described here could be applied to any type of martensitic transformation, and is not limited to the case of zirconia. AFM appears as an extremely powerful and well adapted tool for investigated any kind of phase transformation inducing a surface relief modification. Clues are also given on the experimental limits of the technique, in particular in regards of the precision that may be reached for measuring surface angle relationships. Precise measurements require a precise knowledge of the tip shape and dimensional characteristics.

## Acknowledgments

The authors would like to thank the CLAMS for using the nanoscope. Financial support of the European Union under the GROWTH2000 program, project BIOKER, reference GRD2-2000- 25039. Authors are grateful to G. Thollet and the CLYME for the SEM images of the probe. Acknowledgements are due to H. El Attaoui for providing the samples of the study.

| Lattice Correspondence | Lattice Invariant Shear * | Magnitude of g | Habit plane normal** | Shape strain ** | Shape strain amplitude | Shear component | Volume change |
|---|---|---|---|---|---|---|---|
| ABC 1 | $(011)[0\bar{1}1]$ | 0.0344 | $\begin{bmatrix} -0.9537 \\ 0.0055 \\ 0.3005 \end{bmatrix}$ | $\begin{bmatrix} -0.0026 \\ 0.0028 \\ 0.1640 \end{bmatrix}$ | 0.1640 | 0.1556 | 0.0518 |
| ABC 2 | $(011)[0\bar{1}1]$ | 0.0344 | $\begin{bmatrix} 0.0915 \\ -0.0171 \\ -0.9956 \end{bmatrix}$ | $\begin{bmatrix} 0.1597 \\ -0.0007 \\ -0.0373 \end{bmatrix}$ | 0.1640 | 0.1556 | 0.0518 |
| BCA 1 | $(110)[1\bar{1}0]$ | 0.0344 | $\begin{bmatrix} 0.0034 \\ 0.3935 \\ -0.9193 \end{bmatrix}$ | $\begin{bmatrix} 0.00300 \\ 0.1751 \\ 0.0186 \end{bmatrix}$ | 0.1761 | 0.1683 | 0.0518 |
| BCA 2 | $(110)[1\bar{1}0]$ | 0.0344 | $\begin{bmatrix} -0.0168 \\ -0.9996 \\ -0.0241 \end{bmatrix}$ | $\begin{bmatrix} -0.0004 \\ -0.0558 \\ 0.1670 \end{bmatrix}$ | 0.1761 | 0.1683 | 0.0518 |
| CAB 1 | $(101)[10\bar{1}]$ | 0.0027 | $\begin{bmatrix} 0.3006 \\ -0.9537 \\ -0.0001 \end{bmatrix}$ | $\begin{bmatrix} 0.1640 \\ -0.0026 \\ -0.0002 \end{bmatrix}$ | 0.1640 | 0.1556 | 0.0518 |
| CAB 2 | $(101)[10\bar{1}]$ | 0.0027 | $\begin{bmatrix} -0.9958 \\ 0.0915 \\ 0.0003 \end{bmatrix}$ | $\begin{bmatrix} -0.0373 \\ 0.1597 \\ 0.0001 \end{bmatrix}$ | 0.1640 | 0.1556 | 0.0518 |

TAB. 1: Main features of the martensite plates corresponding to the three correspondences from results of Kelly and Rose [12] and present calculations using Kajiwara phenomenological crystallographic approach [9]. Input parameters: $a_t = 0.5128$ nm, $c_t = 0.5224$, $a_m = 0.5203$, $b_m = 0.5217$, $c_m = 0.5388$, $\beta = 98.91°$, * expression in the lattice axis system of the tetragonal parent phase ** expression in the orthonormal axis system bounded to the tetragonal lattice axis system

| Type 1 | | | Type 2 | | |
|---|---|---|---|---|---|
| ABC1 | $\begin{bmatrix} -0.95 \\ 0 \\ 0.30 \end{bmatrix}_{hpn}$ | $\begin{bmatrix} 0 \\ 0 \\ 0.164 \end{bmatrix}_{ss}$ | ABC2 | $\begin{bmatrix} 0.1 \\ 0 \\ -0.99 \end{bmatrix}_{hpn}$ | $\begin{bmatrix} 0.160 \\ 0 \\ -0.037 \end{bmatrix}_{ss}$ |
| BCA1 | $\begin{bmatrix} 0 \\ 0.39 \\ -0.92 \end{bmatrix}_{hpn}$ | $\begin{bmatrix} 0 \\ 0.175 \\ 0.019 \end{bmatrix}_{ss}$ | BCA2 | $\begin{bmatrix} -0.01 \\ -0.99 \\ -0.03 \end{bmatrix}_{hpn}$ | $\begin{bmatrix} 0 \\ -0.056 \\ 0.167 \end{bmatrix}_{ss}$ |
| CAB1 | $\begin{bmatrix} 0.30 \\ -0.95 \\ 0 \end{bmatrix}_{hpn}$ | $\begin{bmatrix} 0.164 \\ 0 \\ 0 \end{bmatrix}_{ssd}$ | CAB2 | $\begin{bmatrix} -0.99 \\ 0.10 \\ 0 \end{bmatrix}_{hpn}$ | $\begin{bmatrix} -0.037 \\ 0.160 \\ 0 \end{bmatrix}_{ss}$ |

TAB. 2: The two kinds of variants, for all correspondences

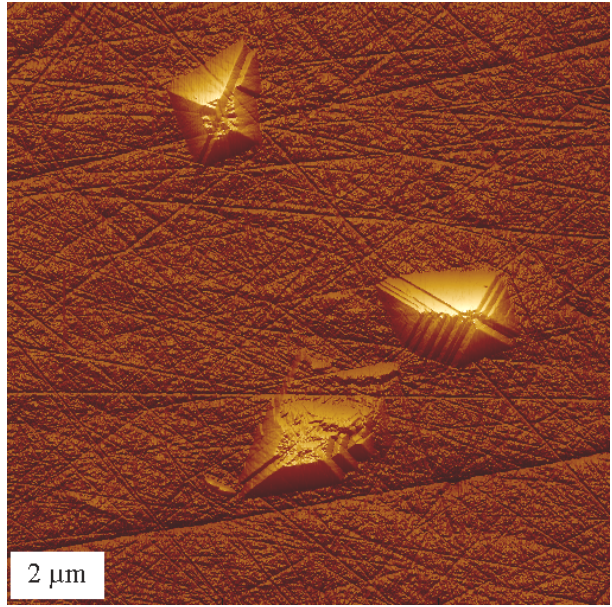

FIG. 1: Transformation induced relief in the first stages of autoclave aging treatment.

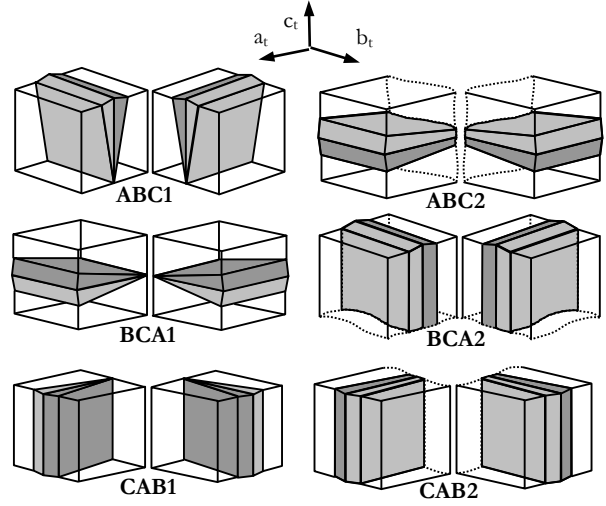

FIG. 2: Self accommodating variants pairs deduced from the various correspondences and from the fourfold symmetry, with the effect on the free surface perpendicular to the junction plane.

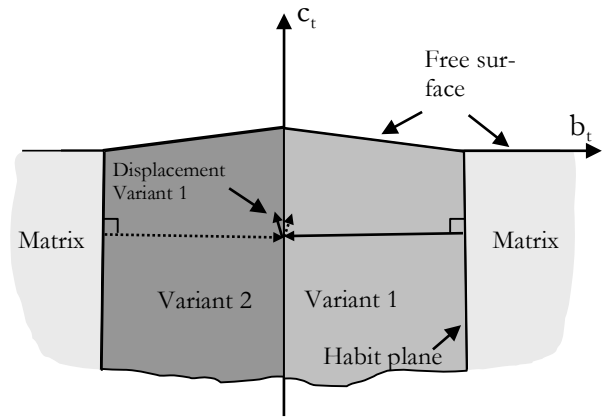

FIG. 3: Behaviour of variants 1 and 2 of correspondence BCA2, close to the free surface.

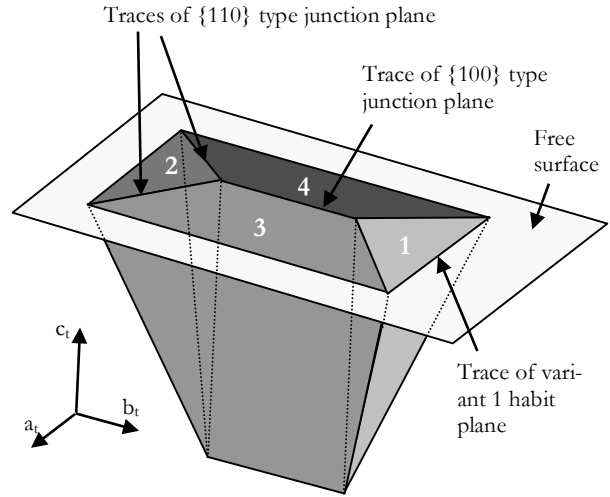

FIG. 4: Simple arrangement of the four variants with correspondence ABC1 when the free surface is $(001)_t$.

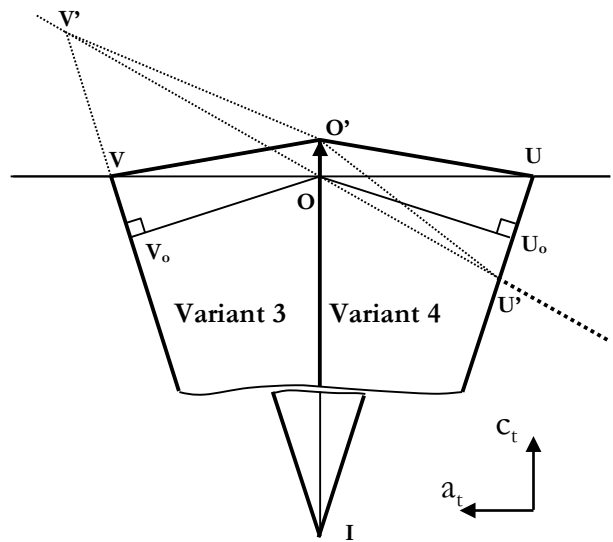

FIG. 5: Spatial arrangement of Fig. 4 in 2D, and effect of the inclination of the free surface in relation to $c_t$.

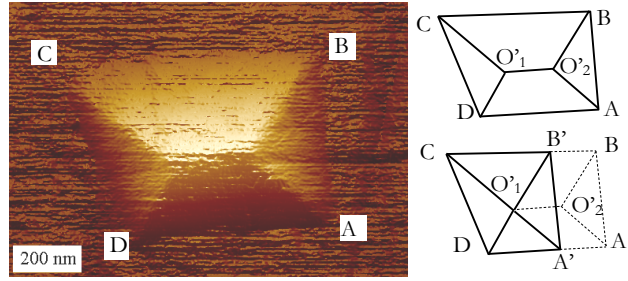

FIG. 6: Elementary arrangement of variants observed by AFM, and extracted traces of the habit and junction planes. The traces can be reduced by translation to a simple *diamond-like* arrangement, where the four $\{110\}_t$ type junction planes traces intersect at a common point ($O'_1$).

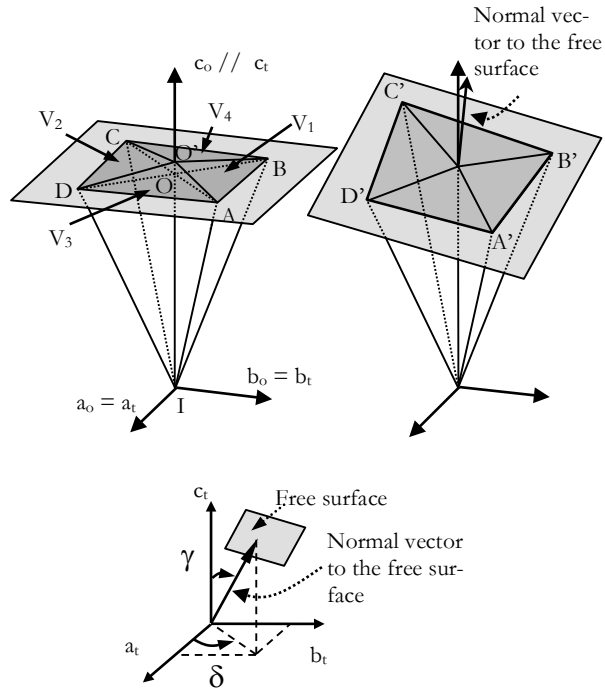

FIG. 7: 3D spatial elementary arrangement of the ABC1 variants, when $c_t$ is perpendicular to the free surface (left), and with an angle between $c_t$ and the free surface (right).

|   | **Experimental observations** |   | **Calculations** (best fit) |
|---|---|---|---|
|   | AFM original image | Experimental trace | g=23 ± 3, d=65 ± 5 |
| 1 | 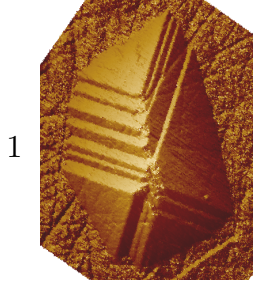 | 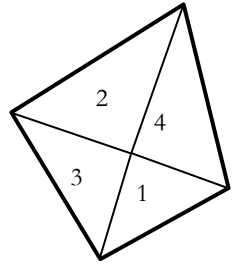 | 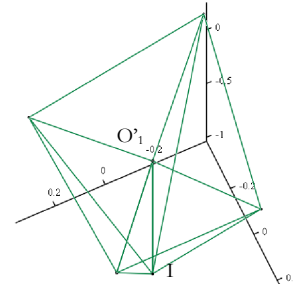 |
|   | (measured) a$_{34}$= 16,3 ± 0,5° a$_{12}$=15,6 ± 0,5° |   | (predicted) a$_{34}$=16,0 ± 0,5° a$_{12}$=15,1 ± 0,7° |
|   |   |   | g=22 ± 3 , d=5 ± 5 |
| 2 | 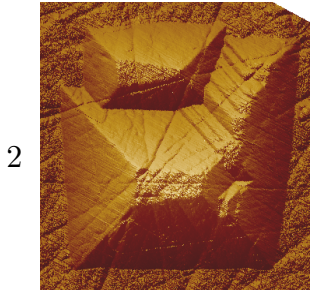 | 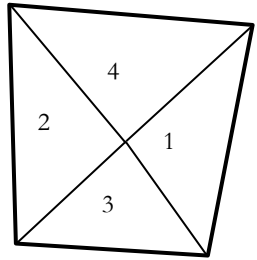 | 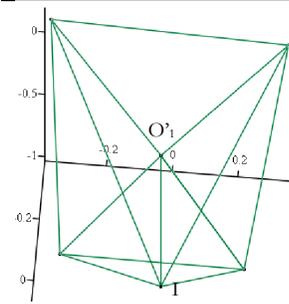 |
|   | a$_{34}$=15,4 ± 0,5° a$_{12}$=15,2 ± 0,5° |   | a$_{34}$=15,1 ± 0,7° a$_{12}$=16,3 ± 0,5° |
|   |   |   | g=27 ± 3, d= -3 ±3 |
| 3 | 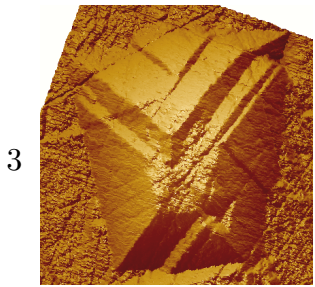 | 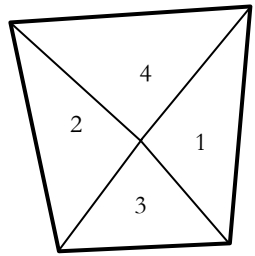 | 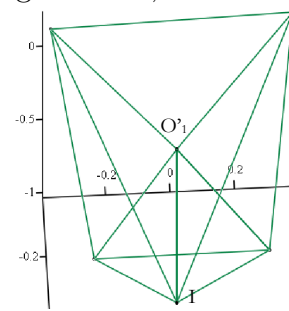 |
|   | a$_{34}$=13,5 ± 0,5° a$_{12}$=14,0 ± 0,5° |   | a$_{34}$=13,8 ± 0,7° a$_{12}$=15,6 ± 0,4° |

FIG. 8: Comparison of the experimental traces and angular relationships between variants and the calculations outputs obtained with the best fit of the model parameters. The best fit is determined the concordance of both the traces and the surface relief angles, when comparing observations and calculations. AFM micrograph and calculated shapes are plotted in a direction normal to the free surface.

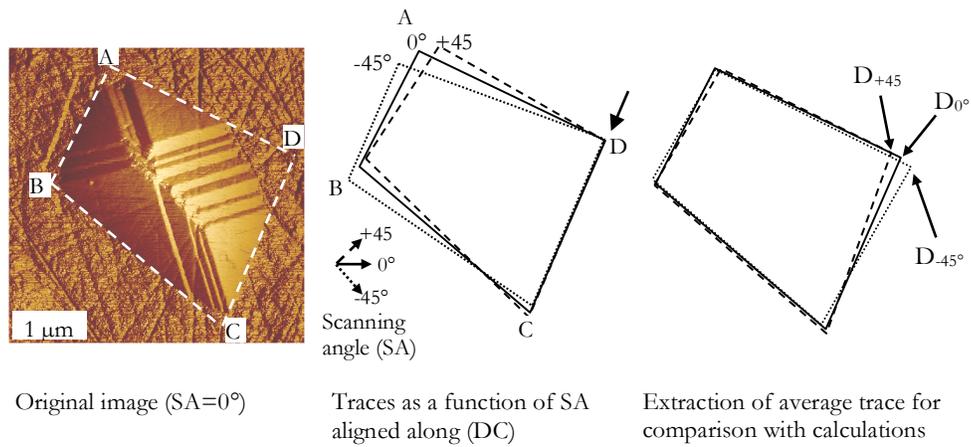

FIG. 9: Assessment of the scanning direction influence on the extracted traces of the variants arrangement.

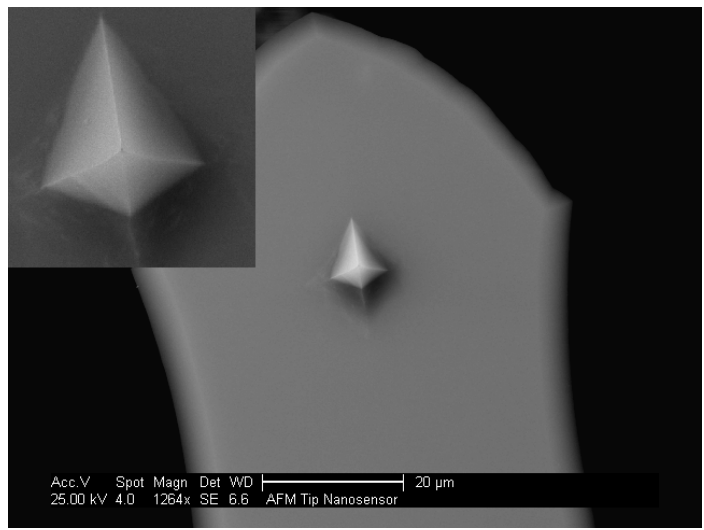

FIG. 10: SEM micrograph of the AFM cantilever and tip, showing the tip shape anisotropy.

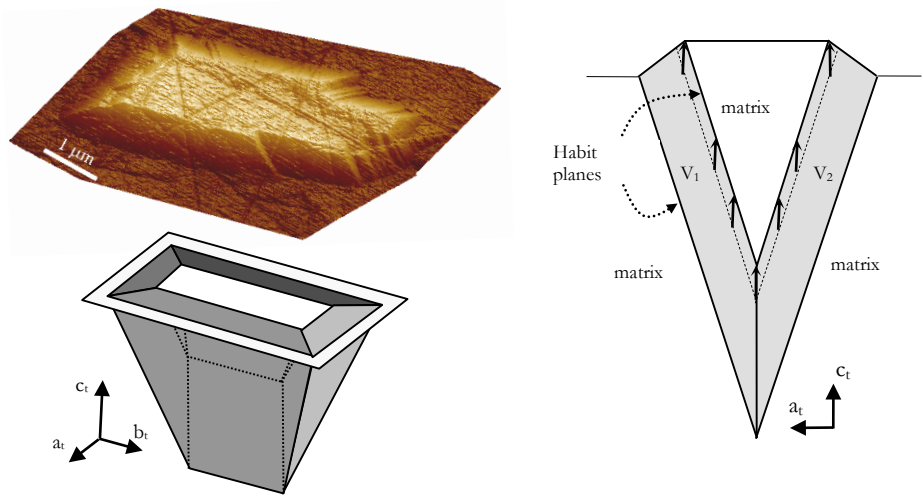

FIG. 11: Partial transformation and corresponding model of the spatial arrangement in the volume and in 2D. All the transformation induced strains are accommodated along $(001)_t$. The displacement is constant along the inner habit planes. The vertical scale of the 2D scheme is dilated for more clarity. The original position of the future inner habit planes is symbolized by the dashed lines.

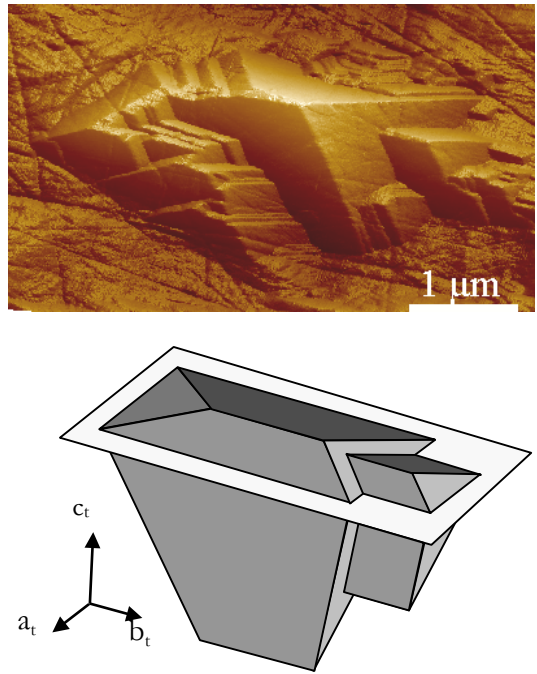

FIG. 12: Secondary variants growth.

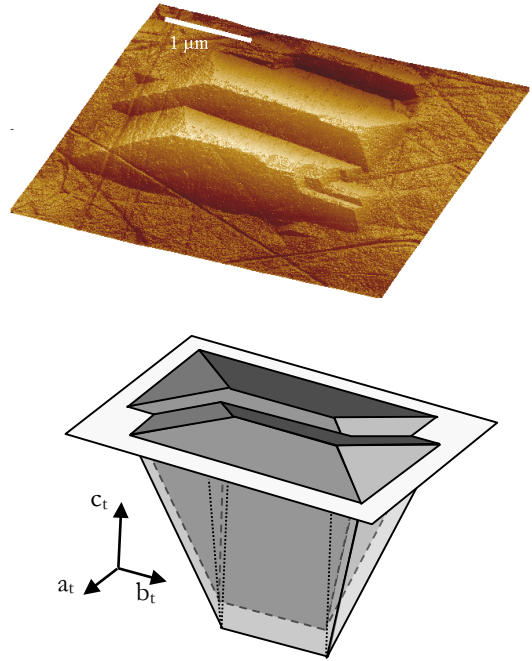

FIG. 13: More complex martensitic features with variants impingement.

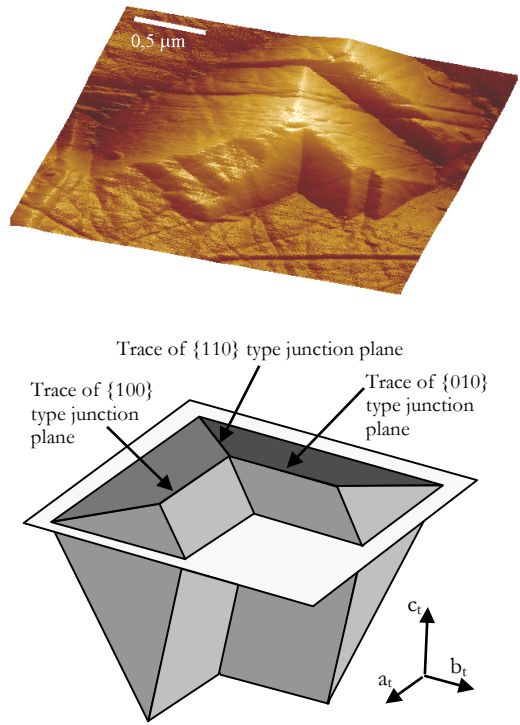

FIG. 14: L-shaped variants, obtained when the three potential junction planes intersect along their common axis.